\documentclass[a4paper,12pt]{article}

\usepackage{indentfirst}
\usepackage{amsmath, amssymb, amscd, amsthm, amsfonts}
\usepackage{mathabx,mathtools}
\usepackage{bm}
\usepackage{graphicx}
\graphicspath{{fig}}
\usepackage{subfigure}
\usepackage{enumitem}
\usepackage{natbib}
\usepackage{url}
\usepackage{hyperref}
\usepackage{multirow}
\usepackage{caption}
\usepackage{color}
\usepackage[margin=1in]{geometry}
\usepackage{authblk}
\usepackage{hhline}
\usepackage{appendix}
\usepackage{soul}
\usepackage{xr}
\usepackage{algorithm,algorithmic}

\allowdisplaybreaks[4]

\makeatletter
\newenvironment{breakablealgorithm}
  {
    \begin{center}
      \refstepcounter{algorithm}
      \hrule height.8pt depth0pt \kern2pt
      \parskip 0pt
      \renewcommand{\caption}[2][\relax]{
        {\raggedright\textbf{\fname@algorithm~\thealgorithm} ##2\par}%
        \ifx\relax##1\relax 
          \addcontentsline{loa}{algorithm}{\protect\numberline{\thealgorithm}##2}%
        \else 
          \addcontentsline{loa}{algorithm}{\protect\numberline{\thealgorithm}##1}%
        \fi
        \kern2pt\hrule\kern2pt
     }
  }
  {
     \kern2pt\hrule\relax
   \end{center}
  }
\makeatother

\theoremstyle{plain}
\newtheorem{theorem}{Theorem}

\newtheorem{corollary}{Corollary}

\newtheorem{remark}{Remark}

\newcommand{\pr}{\mathsf{P}}

\newcommand{\pre}{(\text{pre})}
\newcommand{\opt}{(\text{opt})}
\newcommand{\FDR}{\text{FDR}}
\newcommand{\FDP}{\text{FDP}}
\newcommand{\mFDR}{\text{mFDR}}
\newcommand{\tto}{\text{o}}
\newcommand{\JC}{(\text{JC})}
\newcommand{\e}{(\text{e})}

\newcommand{\calC}{\mathcal{C}}

\newcommand{\calG}{\mathcal{G}}
\newcommand{\calH}{\mathcal{H}}
\newcommand{\calI}{\mathcal{I}}

\newcommand{\calQ}{\mathcal{Q}}
\newcommand{\calR}{\mathcal{R}}
\newcommand{\calS}{\mathcal{S}}
\newcommand{\calT}{\mathcal{T}}

\newcommand{\calX}{\mathcal{X}}
\newcommand{\calY}{\mathcal{Y}}

\newcommand{\bbE}{\mathbb{E}}
\newcommand{\bbI}{\mathbb{I}}
\newcommand{\bbP}{\mathbb{P}}
\newcommand{\bbR}{\mathbb{R}}

\begin{document}

\title{A Unified Framework for Large-Scale Inference of Classification: Error Rate Control and Optimality
}

\author{Yinrui Sun and Yin Xia \\

Department of Statistics and Data Science, Fudan University}

\date{ }

\maketitle

\begin{abstract}\normalsize

Classification is a fundamental task in supervised learning, while achieving valid misclassification rate control remains challenging due to possibly the limited predictive capability of the classifiers or the intrinsic complexity of the classification task.
In this article, we address large-scale multi-class classification problems with general error rate guarantees to enhance algorithmic trustworthiness.
To this end, we first introduce a notion of group-wise classification, which unifies the common class-wise and overall classifications as special cases.
We then develop a unified inference framework for the general group-wise classification that consists of three steps: Pre-classification, Selective $p$-value construction, and large-scale Post-classification decisions (PSP).
Theoretically, PSP is distribution-free and provides valid finite-sample guarantees for controlling general group-wise false decision rates at target levels.
To show the power of PSP, we demonstrate that the step of post-classification decisions never degrades the power of pre-classification, provided that pre-classification has been sufficiently powerful to meet the target error levels.
We further establish general power optimality theories for PSP from both non-asymptotic and asymptotic perspectives.
Numerical results in both simulations and real data analysis validate the performance of the proposed PSP approach.
In addition, we introduce an ePSP algorithm that integrates the idea of PSP with selective $e$-values.
Finally, extensions of PSP are shown to demonstrate its feasibility and power in broader applications. 

\end{abstract}

Keywords: Multi-class classification; Algorithmic trustworthiness; Conformal inference; Selective inference; False decision rate control.

\newpage

\baselineskip=20pt

\section{Introduction}\label{sec:intro}

Classification is a fundamental task in supervised learning that aims to assign data points to different categories based on the observed features, and it serves as a crucial role in data-driven predictive analysis and decision-making.
Over the past several decades, a wide range of classification algorithms have been developed, including logistic regression, discriminant analysis, nearest neighbors, random forests, support vector machines, neural networks, among others.
These classifiers have been widely applied across various domains, such as medical diagnosis, image recognition, fraud detection, and automatic data labelling.
The emergence of big data with increasingly complex patterns has further underscored the need to achieve trustworthy classification.

Nonetheless, ensuring trustworthiness in modern classifiers remains a significant challenge, particularly in terms of error controllability and class fairness.
First, classification accuracy can be affected by multiple factors, including algorithm selection, hyperparameter tuning, model training, as well as external factors such as limited training sample sizes and the intrinsic complexity of the classification task. 
These uncertainties make it challenging to maintain misclassification errors below a pre-specified target threshold.
The challenge is particularly pronounced in risk-sensitive applications, where even minimal errors can lead to severe risks and a low error level is often highly desirable. 
Second, fairness in classification may be compromised for classes that are inherently more complex to classify or are less represented in the training data.
In such cases, classifiers may exhibit biases that favor the dominant classes and underperform on underrepresented or complex ones. 
This issue is especially critical in scenarios where the class-wise predictions are of primary interest.
Given these challenges, developing classification algorithms with controllable error and fairness guarantees is essential for ensuring trustworthy predictive analysis and decision-making in practical applications.

\begin{figure}[t]
\centering
\includegraphics[scale=0.3]{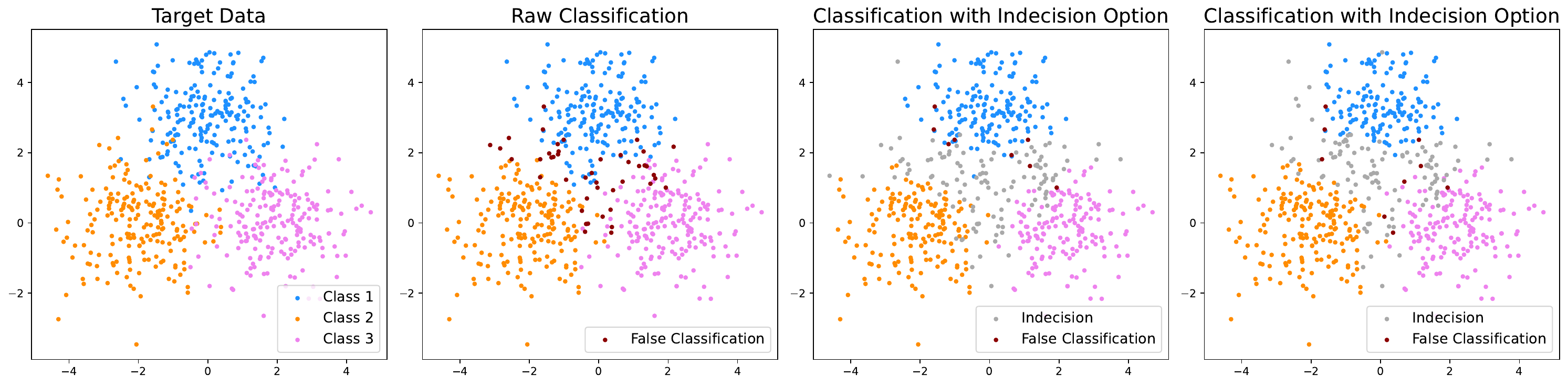}
\caption{Illustration for classification with indecision option. The left panel presents the data with 3 classes. The middle-left panel shows the classification results by LightGBM. The two right panels show the classification results with indecisions by the proposed PSP approach in Algorithm \ref{alg:method} with overall and class-wise error rate control, respectively.}
\label{fig:illustration}
\end{figure}

In this article, we aim to address the large-scale classification problems with general error rate guarantees, including both class-wise and overall error control.
Specifically, in the classification tasks with $K$ classes, suppose that covariate features $X_1,\cdots,X_m \in \calX$ are observed for $m$ subjects, where $\calX$ represents the feature space, and our objective is to simultaneously predict the corresponding class labels $Y_1,\cdots, Y_m \in [K] := \{1,2,\cdots, K\}$ while controlling the general error rates as detailed in Section \ref{sec:problem}.
To ensure the trustworthiness of classification, we introduce an additional indecision option alongside the $K$ class labels.
This allows the final label predictions $\widehat{Y}_1,\cdots, \widehat{Y}_m$ to take values in $\{0\} \cup [K]$, where the prediction of $0$ indicates the indecision.
The indecision option serves as a rejection mechanism and enables the classifier to abstain from making a prediction when there is low confidence, which may arise from either the classifier's  limited predictive capability or the intrinsic ambiguity of certain subjects.
By involving the indecisions, the classification potentially avoids unreliable predictions, thereby reducing the false classifications.
Moreover, beyond improving the classification confidence, the indecision mechanism also facilitates the automatic identification of ambiguous subjects that may require further assessment such as human-intervened decisions.
Such indecision option to reject classification is illustrated in Figure \ref{fig:illustration}.

\subsection{Related Works}\label{sec:intro_review}

Classification with indecision options, also referred to as classification with rejection, addresses the trade-off between misclassification and indecision risks \citep{cho1970optimum}.
By assigning pre-specified risks for misclassifications and indecisions, most existing works focus on total risk minimization to learn classification and rejection rules \citep{bartlett2008classification,zhang2018reject,ni2019calibration,
mao2024predictor,mao2024theoretically,mohri2024learning}.
In addition, \citet{ndaoud2024ask} develops theoretical results for classification with indecisions under either fixed rejection or misclassification rates.
Despite these contributions, there remains a lack of inference methodologies for classifications with valid error rate control.

Quantifying uncertainty and evaluating confidence in classification present significant challenges in modern data analysis due to the model complexity and algorithmic learning errors.
Under parametric modelling for classification probability functions, the idea of bias correction \citep{javanmard2014confidence,zhang2014confidence,van2014on,
chernozhukov2018double} has been employed to provide uncertainty quantification for estimated probability functions \citep{guo2021inference,hou2023surrogate,zhou2025doubly}.
However, these approaches highly rely on the specific parametric model structures and large-sample asymptotics, which may fail in scenarios with complex data structures or limited sample sizes.

In contrast, conformal prediction approaches \citep{vovk2005algorithmic,angelopoulos2023conformal} provide distribution-free predictive analysis with confidence guarantees.
Their advantages in model-agnostic analysis and capability for uncertainty quantification facilitate the broad applications in regressions \citep{lei2018distribution,tibshirani2019conformal,
romano2019conformalized,barber2021predictive,lei2021conformal,dunn2023distribution,
lee2023distribution,cauchois2024robust,yang2025selection,liang2025conformal} and classifications \citep{lei2014classification,sadinle2019least,romano2020classification,
guan2022prediction,gazin2025powerful,sesia2025adaptive,bortolotti2025noise}, where prediction sets are constructed with controlled miscoverage rates.
Moreover, the ideas of conformal inference have been extended to selective inference \citep{bao2024selective,bao2024cap,jin2025confidence,
gazin2025selecting,sale2025online} and multiple testing \citep{mary2022semi,bates2023testing,jin2023selection,jin2023model,
bashari2023derandomized,gao2023simultaneous,marandon2024adaptive,liang2024integrative,
bai2024optimized,gao2025adaptive,lee2025full,lee2025selection}.

Several recent works address confidence evaluation for classification with indecision options.
\citet{gang2024locally} studies large-scale binary classification under Gaussian models and proposes consistent estimation procedure to achieve asymptotic class-wise error rate control.
\citet{rava2024burden} introduces a conformal inference approach for binary classification with error control targeted at one specific class, a problem also studied under conformal selection \citep{jin2023selection}, with additional consideration for covariate subgroups.
\citet{zhao2023controlling} studies overall error rate control in binary classification.
\citet{gazin2025selecting} addresses informative conformal prediction set construction with overall error rate control, which covers multi-class classification with indecisions by defining informative sets as those containing only a single class label.
\citet{rava2025burden}, a recent updated version of \citet{rava2024burden}, achieves class-wise error control for both classes in binary classification by considering the decision rules in \citet{gang2024locally,zhao2023controlling}.
Despite these advances, however, developing classification algorithms for multi-class problems with general target error rate control remains an ongoing challenge.

\subsection{Contributions}

In this article, we aim to develop valid and powerful large-scale classification algorithm with indecision options to address multi-class classification problems with general error rate guarantees.
Notably, we introduce a notion of group-wise classification, which unifies the overall and class-wise classifications along with the corresponding error rates.
Methodologically, we propose a novel algorithmic framework for general group-wise classification that consists of three steps: Pre-classification, Selective $p$-value construction, and large-scale Post-classification decisions (PSP).
First, the pre-classification step conducts preliminary classification $\widehat{Y}_j^{\pre} \in [K]$ for the target covariate $X_j$, $j\in[m]$.
Next, the selective $p$-value construction step quantifies the uncertainties and provides the confidence levels for the pre-classification results.
Finally, the step of large-scale post-classification decisions incorporates the constructed selective $p$-values into the proposed large-scale selective inference procedures to yield the final classification $\widehat{Y}_j \in \{0, \widehat{Y}_j^{\pre} \}$ for each $j\in[m]$.

Under the proposed PSP framework, we establish valid finite-sample guarantees for controlling general group-wise false decision error rates, which accommodate both overall and class-wise error control.
To show the power of PSP approach, we demonstrate that post-classification decisions never degrade the accuracy of pre-classification if it has been sufficiently powerful to meet the target error levels.
We further establish general optimality theories for the power of PSP by first deriving the oracle optimal decisions and then showing the approximations of PSP to the oracle from both non-asymptotic and asymptotic perspectives.
Numerical studies in simulations and real data analysis validate the performance of the proposed PSP approach.
In addition, we develop an ePSP algorithm that integrates the idea of PSP with selective $e$-values, and establish the error rate guarantees as well as its connections to PSP.
Finally, we show the extensions of PSP in broader applications and demonstrate its validity and power in problems of subject selection and informative prediction set construction. 

Overall, this article makes novel contributions to both methodology and theory.
First, to the best of our knowledge, the proposed PSP framework is the first to accommodate both the overall and class-wise error rate control for general multi-class classification problems.
This advancement address a significant methodological gap in the existing literature.
Second, the validity of error rate control is rigorously guaranteed in finite-sample settings, and the PSP algorithm is distribution-free and supports the flexible integration of various machine learning classifiers in the implementation.
Third, the establishes optimality theories for PSP contributes to the power analysis of model-free inference procedures.
Moreover, due to the incorporation of the idea of conformal inference in constructing the selective $p$-values, the derived optimality results also shed light on the power of conformal approaches in inference problems.
Fourth, the extensions of the PSP framework to the problems of subject selection and informative prediction set construction demonstrate its feasibility and power in broader applications.

\subsection{Article Organization}

The remainder of the article is organized as follows.
Section \ref{sec:problem} formulates the problem setup and introduces a general framework of group-wise classification which unifies overall and class-wise classifications as special cases.
Section \ref{sec:method} presents the PSP algorithm for general group-wise classification.
The steps of pre-classification, selective $p$-value construction and large-scale post-classification decisions are described in Sections \ref{sec:method_preclass}-\ref{sec:method_inference}, respectively.
Section \ref{sec:theory_error} establishes the validity of PSP in error rate control, and Section \ref{sec:theory_power} develops theoretical results for power analysis.
Sections \ref{sec:simulation}-\ref{sec:realdata} present simulation studies and real data analysis.
Section \ref{sec:evalue} introduces the ePSP algorithm that integrates the PSP framework with selective $e$-values.
Extensions of PSP and its broader applications are shown in Section \ref{sec:extend}.
Additional results and technical proofs are included in the supplementary material.

\section{Problem Formulation}\label{sec:problem}

In multi-class classification problems with $K$ classes, we consider the setting where covariates $X_1,\cdots,X_m \in \calX$ are observed for $m$ subjects, and the corresponding class labels $Y_1,\cdots, Y_m \in [K]$ are to be predicted simultaneously by leveraging an additional available labelled hold-out dataset $(X_{m+i}, Y_{m+i}) \in \calX \times [K], i\in[n]$.
To achieve trustworthy classification with error rate control, as introduced in Section \ref{sec:intro}, we adopt the rejection mechanism that allows for indecision options for classification, yielding the final label predictions $\widehat{Y}_1,\cdots, \widehat{Y}_m \in \{0\} \cup [K]$, where $0$ indicates the rejection to make a classification.
Such a rejection mechanism enables the algorithm to abstain from making decisions in cases of high uncertainty, that may arise from the limited predictive capacity of classifier or the inherent ambiguity of certain subjects.
Therefore, it potentially enhances the classification confidence and then ensures that the misclassification rate can be controlled below the target error level.
Moreover, the subjects with indecisions can rather be identified as ambiguous cases that require further human-intervened evaluation and decision-making.

Building on the indecision options for classification, following the idea of false discovery rate \citep{benjamini1995controlling} for multiple testing problems, we propose to control the following overall false decision rate (FDR), defined as the expectation of the false decision proportion (FDP):
\begin{equation}\label{eq:FDR}
\FDR_{\tto} = \bbE \FDP_{\tto} , ~ \FDP_{\tto} = \frac{ \sum_{j\in[m]} \bbI\left( \widehat{Y}_j \neq 0, \widehat{Y}_j \neq Y_j \right) }{1 \vee \sum_{j\in[m]} \bbI\left( \widehat{Y}_j \neq 0 \right)} ,
\end{equation}
which is also referred to as false selection rate \citep{gang2024locally}.
While the $\FDR_{\tto}$ provides the measure of overall classification error among the decisions, however, it does not fully account for the class-wise performance of the classification.
Thus, in addition to the overall $\FDR_{\tto}$ in \eqref{eq:FDR}, we also consider the class-wise FDR for error evaluation within individual classes:
\begin{equation}\label{eq:FDR_k}
\FDR_k = \bbE \FDP_k , ~ \FDP_k = \frac{ \sum_{j\in[m]} \bbI\left( \widehat{Y}_j = k,  Y_j \neq k \right) }{1 \vee \sum_{j\in[m]} \bbI\left( \widehat{Y}_j = k \right)} , ~ k \in [K] .
\end{equation}

To further generalize the error rate control, we unify the overall $\FDR_{\tto}$ and class-wise $\FDR_k,k\in[K]$ above by introducing a notion of group-wise false decision rate.
Specifically, let $\calG_1, \cdots, \calG_G \subset [K]$ form a disjoint partition of the class set $[K]$ such that $[K] = \bigcup_{g\in[G]} \calG_g$ and $\calG_g \bigcap \calG_{g^\prime} = \phi$ for $g \neq g^\prime$, and we define the group-wise FDR as
\begin{equation}\label{eq:FDR_g}
\FDR_{ \calG_g } = \bbE \FDP_{ \calG_g }, ~  \FDP_{ \calG_g } = \frac{ \sum_{j\in[m]} \bbI\left( \widehat{Y}_j \in \calG_g, \widehat{Y}_j \neq Y_j \right) }{1 \vee \sum_{j\in[m]} \bbI\left( \widehat{Y}_j \in \calG_g \right)} , ~ g \in [G] .
\end{equation}
This group-wise notion links to both overall and class-wise error rates. Specifically, when $G = 1$, we have $\calG_1 = [K]$ and $\FDR_{ \calG_1 } = \FDR_{\tto}$ in \eqref{eq:FDR}, whereas when $G = K$ with $\calG_k = \{k\}$, we have $\FDR_{ \calG_k } = \FDR_k$ in \eqref{eq:FDR_k} for $k\in[K]$.
Thus, the introduced group-wise $\FDR_{ \calG_g },g\in[G]$ provides a unified measure for evaluating the classification errors, and includes the overall and class-wise error rates as special cases.

In this article, given a predetermined general partition $\{ \calG_g \}_{ g\in[G] }$ that may be specified according to user’s interests and targets, our goal is to develop a powerful large-scale classification algorithm with valid control of $\FDR_{ \calG_g },g\in[G]$ at target error levels $\alpha_{\calG_g},g\in[G]$.
Throughout, we assume the data $(X_j, Y_j)_{ j\in[m+n] }$ are independent and identically distributed (i.i.d.) from a common distribution. 
Although a milder exchangeability condition is sometimes imposed in conformal inference literature, the i.i.d setting remains standard in statistical analysis and accommodates a wide range of problems.

\section{Methodology}\label{sec:method}

In this section, to address the large-scale multi-class classification problems with valid group-wise FDR control, we develop a unified framework for group-wise classification with indecisions that consists of three steps: Pre-classification, Selective $p$-value construction, and large-scale Post-classification decisions (PSP). 
First, in the pre-classification step, the covariates of interest $X_1,\cdots,X_m$ are initially labelled by a pilot classifier, yielding preliminary labels $\widehat{Y}_1^{\pre},\cdots,\widehat{Y}_m^{\pre} \in [K]$.
Second, selective $p$-values are constructed for each subject by using the labelled hold-out data $(X_{m+1}, Y_{m+1}),\cdots, (X_{m+n}, Y_{m+n})$, which provide uncertainty quantification for the pre-classification results in the first step.
Finally, the constructed selective $p$-values are incorporated into the proposed large-scale post-classification inference procedures to make final decisions $\widehat{Y}_j \in \{ 0, \widehat{Y}_j^{\pre} \}$ for $j\in[m]$.

The developed PSP approach for general group-wise classification is summarized in Algorithm \ref{alg:method}, and will be detailed in Sections \ref{sec:method_preclass}-\ref{sec:method_inference}.
The specific implementations of PSP for overall and class-wise classifications are provided in Section \ref{sec:alg_overall_classwise} of the supplement as special cases.

\begin{breakablealgorithm}
\caption{PSP for General Group-Wise Classification}
\label{alg:method}
\begin{algorithmic}
~

\textbf{Inputs:} Target data $X_1,\cdots,X_m$; labelled hold-out data  $(X_{m+1}, Y_{m+1}), \cdots, (X_{m+n}, Y_{m+n})$; partition groups $\calG_1, \cdots, \calG_G$; pre-classification algorithm \texttt{PreClass}; score function $\mu = \left( \mu_1,\cdots,\mu_K \right) : \mathcal{X} \rightarrow \mathbb{R}^{K}$; target error levels $\alpha_{ \calG_g } \in (0,1)$, $g\in[G]$.

\textbf{PSP:}
\begin{enumerate}[nosep]

\item \textbf{Pre-classification.}

\begin{enumerate}[nosep]
\item Let $\widehat{Y}_j^{\pre} = \texttt{PreClass}\left(  X_j, U_j \right) \in [K], j\in[m+n]$ be the pre-classification results for target data and labelled hold-out data, where $U_1,\cdots,U_{m+n}$ represent the possible randomness in \texttt{PreClass}.

\item For $k\in[K]$, define the pre-classification sets for target data and labelled hold-out data: 
$$
\calS_{k} = \left\{ j \in [m]: \widehat{Y}_j^{\pre} = k \right\} \text{ and } \widebar{\calS}_{k} = \left\{ i \in [n]: \widehat{Y}_{m+i}^{\pre} = k \right\}.
$$

\end{enumerate}

\item \textbf{Selective $p$-value construction.}

For $g \in [G]$, do:

\quad  For $k\in \calG_g$, define the selective $p$-values:
$$
p_{j} = \frac{1}{ 1 + \sum_{ k^\prime \in \calG_g } | \widebar{\calR}_{ k^\prime } | } \left( 1 + \sum_{ k^\prime \in \calG_g } \sum_{ i \in \widebar{\calR}_{k^\prime} } \bbI \left( \mu_{k^\prime} (X_{m+i}) \geq \mu_k( X_j ) \right) \right) , ~ j \in \calS_k,
$$
\quad where $\widebar{\calR}_{ k^\prime } = \left\{ i \in \widebar{\calS}_{ k^\prime } : Y_{m+i} \neq k^\prime \right\}$.

\item \textbf{Large-scale post-classification decisions.}

For $g \in [G]$, do:

\begin{enumerate}[nosep]
\item Let $p_{(\calG_g,1)} \leq \cdots \leq p_{(\calG_g, \sum_{ k\in \calG_g } |\calS_k|)}$ denote the sorted values of $\left( p_j \right)_{ j \in \bigcup_{ k\in \calG_g } \calS_k }$ in ascending order.

\item Define $\widehat{\theta}_{ \calR, \calG_g } = \left( 1 + \sum_{ k\in \calG_g } | \widebar{\calR}_k | \right) / \left( 1 + \sum_{ k\in \calG_g } | \widebar{\calS}_k | \right)$.

\item Data-driven threshold:
$$
\widehat{T}_{ \calG_g } = p_{( \calG_g , \widehat{l}_{ \calG_g } )}, ~ \widehat{l}_{ \calG_g } = \max\left\{ l \in [ \sum_{ k\in \calG_g } |\calS_k| ]: p_{( \calG_g,l )} \leq \frac{ l \alpha_{ \calG_g } }{ \widehat{\theta}_{ \calR, \calG_g } \sum_{ k\in \calG_g } |\calS_k| } \right\} .
$$
If such $\widehat{l}_{ \calG_g }$ doesn't exist, set $\widehat{T}_{ \calG_g } = 0$.

\item Post-classification decisions: 
$
\widehat{Y}_j = \widehat{Y}_j^{\pre} \bbI ( p_j \leq \widehat{T}_{ \calG_g } )$ for $j \in \bigcup_{ k\in \calG_g } \calS_k$.

\end{enumerate}

\end{enumerate}

\textbf{Outputs:} $\widehat{Y}_j \in \left\{ 0 \right\} \bigcup [K]$ for  $j\in[m]$.

\end{algorithmic}
\end{breakablealgorithm}

\subsection{Pre-Classification}\label{sec:method_preclass}

The first step of Algorithm \ref{alg:method} performs pre-classification that assigns the covariates of interest $X_1,\cdots,X_m$ to predicted class labels $\widehat{Y}_1^{\pre},\cdots,\widehat{Y}_m^{\pre} \in [K]$.
Given a pilot classifier $\mu^{\pre}$, the pre-classification rule for $X_j$'s is generally expressed as $\widehat{Y}_j^{\pre} = \texttt{PreClass}(\mu^{\pre}, X_j, U_j)$, where $U_j$ captures the possible additional randomness in pre-classification procedure.
Several examples of classifiers and the corresponding classification rules are outlined below:
\begin{itemize}[nosep]
\item $\mu^{\pre} : \calX \rightarrow [K]$ is a hard-classifier, and the pre-classification rule is given by $\widehat{Y}_j^{\pre} = \mu^{\pre}(X_j)$.

\item $\mu^{\pre} =  (\mu_1^{\pre},\cdots,\mu_K^{\pre}): \calX \rightarrow \bbR^K$ is a soft-classifier, where each component $\mu_k^{\pre}$ represents the confidence level for class $k$, and the pre-classification rule is given by $\widehat{Y}_j^{\pre} = \arg\max_{ k\in[K] } \mu^{\pre}_k(X_j)$.
In cases where multiple classes attain the maximum value, $\widehat{Y}_j^{\pre}$ can be uniformly sampled from the set $\arg\max_{ k\in[K] } \mu^{\pre}_k(X_j)$, where the sampling randomness is encoded by $U_j$.

\end{itemize}
The two examples are commonly used in practice and are provided here for illustration, while the proposed methodology remains general and is not restricted to these cases.

Following the pre-classification step for target covariates, we conduct the same pre-classification procedure for the covariates $X_{m+1},\cdots,X_{m+n}$ in the hold-out dataset. 
This yields the predicted class labels $\widehat{Y}_{m+i}^{\pre} := \texttt{PreClass}(\mu^{\pre}, X_{m+i}, U_{m+i}) \in [K]$ for $i\in[n]$, where $U_{m+1},\cdots,U_{m+n}$ represents possible additional randomness.
Given the pre-classified labels $\widehat{Y}_1^{\pre},\cdots,\widehat{Y}_{m+n}^{\pre}$, 
we define the pre-classification sets for the target data and hold-out data respectively for each $k\in[K]$:
\begin{equation}\label{eq:preclass_set}
\calS_{k} = \left\{ j \in [m]: \widehat{Y}_j^{\pre} = k \right\} \text{ and } \widebar{\calS}_{k} = \left\{ i \in [n]: \widehat{Y}_{m+i}^{\pre} = k \right\} .
\end{equation}
These sets serve as pilot label predictions similar to standard classification procedures in practical data analysis.
While generally the pre-classification step does not provide any valid misclassification control, the subsequent two steps introduce inference procedures to quantify uncertainty and refine the pre-classification results.

It is worthy noting that we impose no restrictions on the choice of pilot classifier $\mu^{\pre}$ or the classification rule $\texttt{PreClass}$, thus providing broad flexibility for various classification problems and classifiers.
The only requirement is that the data $(X_j, Y_j, U_j),j\in[m+n]$ maintain independence and identical distribution, and are independent of the pilot classifier $\mu^{\pre}$.
This requirement is mild, and the classifier $\mu^{\pre}$ can be obtained in various ways, such as training a classifier or fine-tuning a pre-trained classifier on an additional hold-out dataset, as in the splitting conformal approach \citep{papadopoulos2002inductive,lei2018distribution}, or directly using an external pre-trained classifier like a large foundation model.
Given the independence structure, $\mu^{\pre}$ is treated as an deterministic function throughout the article, and the pre-classification is abbreviated as $\widehat{Y}_j^{\pre} := \texttt{PreClass}(X_j, U_j)$ for $j\in[m+n]$.

\begin{remark}\label{remark:preclass}
The flexibility of the pre-classification step lies not only in the choice of \texttt{PreClass} and $\mu^{\pre}$, but also in its adaptability to user’s specific targets and interests. 
For example, if only a subset of groups $\{ \calG_g \}_{g \in \calI}$ for some $\calI \subset [G]$ is of interest, the output of \texttt{PreClass} can be restricted to the classes $\bigcup_{g \in \calI} \calG_g \subset [K]$ in practice.  
Such accommodation facilities the extension of PSP to broader applications as shown in Section \ref{sec:extend}.
\end{remark}

\subsection{Selective $p$-value Construction}\label{sec:method_pvalue}

Based on the pre-classification results, the second step constructs selective $p$-values for each subject $j\in[m]$ to evaluate the evidence for the pre-classification results by employing a score function $\mu = \left( \mu_1,\cdots,\mu_K \right) : \mathcal{X} \rightarrow \mathbb{R}^{K}$, where each component $\mu_k$ represents the confidence level for class $k$.
Similarly to the pilot classifier $\mu^{\pre}$ in pre-classification step, no restriction is imposed on the score $\mu$ except that it is assumed to be independent of $\left( (X_1, Y_1), \cdots, (X_{m+n}, Y_{m+n}) \right)$ and $\left(U_1,\cdots,U_{m+n} \right)$, and such a function can be obtained via an external model or a model trained on another hold-out dataset.
Due to this independence, $\mu$ is viewed as a deterministic function throughout the article.

The construction of the selective $p$-values now proceeds as follows.
Based on the predetermined groups $\calG_1,\cdots,\calG_G$, we partition the sample index set $[m]$ of target data by leveraging the pre-classification sets obtained in Section \ref{sec:method_preclass}: 
$
[m] = \bigcup_{ g\in[G] } \bigcup_{ k\in\calG_g } \calS_k .
$
Then, given the score function $\mu$, for each group $g\in [G]$ and for each class $k\in\calG_g$, the selective $p$-values are defined as:
\begin{equation}\label{eq:pvalue}
p_{j} = \frac{1}{ 1 + \sum_{ k^\prime \in \calG_g } | \widebar{\calR}_{ k^\prime } | } \left( 1 + \sum_{ k^\prime \in \calG_g } \sum_{ i \in \widebar{\calR}_{k^\prime} } \bbI \left( \mu_{k^\prime} (X_{m+i}) \geq \mu_k( X_j ) \right) \right) , ~ j \in \calS_k,
\end{equation}
where 
$$
\widebar{\calR}_{ k^\prime } := \left\{ i \in \widebar{\calS}_{ k^\prime } : Y_{m+i} \neq k^\prime \right\} \text{ for } k^\prime\in[K] .
$$
Such a construction follows the idea of conformal $p$-value construction \citep{mary2022semi,bates2023testing,jin2023selection,marandon2024adaptive}, while the data-dependent selective nature of \eqref{eq:pvalue}, both in the selection of subjects $j \in \calS_k$ and the selection of scores $\{ \mu_{k^\prime}(X_{m+i}) \}_{i \in \widebar{\calR}_{k^\prime}, k^\prime \in \calG_g}$ based on the pre-classification results in the first step, distinguishes it from standard conformal $p$-values and exhibits its novelty.

It is worthy noting that although the classification task is not a straightforward hypothesis testing problem, the constructed selective $p_j$ in \eqref{eq:pvalue} can be interpreted as an evidence measure against the conceptualized null ``$\calH_{0,j}:X_j$ does not belong to the class $k$, given that $X_j$ is assigned to the class $k$ in group $\calG_g$ in the pre-classification step'', for $j\in \calS_k,  k\in\calG_g, g\in[G]$.
It is clear that for group $\calG_g$, the corresponding $p$-values are valued in $(0,1)$, and a smaller value of $p_j$ implies stronger confidence against the $\calH_{0,j}$, thereby suggesting greater reliability of the pre-assigned class label.

\subsection{Large-Scale Post-Classification Decisions}\label{sec:method_inference}

The final step conducts large-scale post-classification decisions $\widehat{Y}_j \in \{ 0, \widehat{Y}_j^{\pre} \}$ simultaneously for $j\in[m]$.
Let $\{ \alpha_{ \calG_g } \}_{ g\in[G] }$ denote the predetermined target error levels. 
To provide the group-wise $\FDR_{ \calG_g }$ control at the level $\alpha_{ \calG_g }$, for each $g\in[G]$, we propose the following group-wise post-classification decisions based on the $p$-values constructed in \eqref{eq:pvalue}:
\begin{equation}\label{eq:postclass_decision}
\widehat{Y}_j = \widehat{Y}_j^{\pre} \bbI \left( p_j \leq \widehat{T}_{ \calG_g } \right), ~ j \in \bigcup_{ k\in\calG_g } \calS_k ,
\end{equation}
where $\widehat{T}_{ \calG_g }$ is a data-driven group-wise threshold for decision-making.
To determine $\widehat{T}_{\calG_g}$, we develop a post-classification selective inference procedure that generalizes the Benjamini-Hochberg \citep[BH,][]{benjamini1995controlling} method.
Specifically, the data-driven procedure for determining the threshold is proposed as: \begin{equation}\label{eq:threshold}
\widehat{T}_{ \calG_g } = p_{( \calG_g, \widehat{l}_{\calG_g} )}, \text{ with } \widehat{l}_{\calG_g} = \max\left\{ l \in [ \sum_{ k\in \calG_g } |\calS_k| ]: p_{( \calG_g, l)} \leq \frac{ l \alpha_{ \calG_g } }{ \widehat{\theta}_{ \calR, \calG_g } \sum_{ k\in \calG_g } |\calS_k| } \right\} ,
\end{equation}
where $p_{( \calG_g, 1 )} \leq \cdots \leq p_{( \calG_g, \sum_{ k\in \calG_g } |\calS_k|)}$ denote the sorted values of $\left( p_j \right)_{ j\in \bigcup_{ k \in \calG_g } \calS_k }$ in ascending order, and
\begin{equation}\label{eq:proportion_RGg_est}
\widehat{\theta}_{ \calR, \calG_g } := \frac{ 1 + \sum_{ k\in \calG_g } | \widebar{\calR}_k | }{ 1 + \sum_{ k\in \calG_g } | \widebar{\calS}_k | } .
\end{equation}
If $\widehat{l}_{ \calG_g }$ in procedure \eqref{eq:threshold} does not exist, we set $\widehat{T}_{ \calG_g } = 0$, i.e., $\widehat{Y}_j = 0$ for all $j \in \bigcup_{ k\in\calG_g } \calS_k$.

To understand the role of $\widehat{\theta}_{ \calR, \calG_g }$, note that it empirically approximates the quantity: 
\begin{equation}\label{eq:proportion_RGg}
\theta_{ \calR, \calG_g } := \mathbb{P} \left( \widehat{Y}_j^{\pre} \neq Y_j \mid j \in \bigcup_{ k \in \calG_g } \calS_k \right) = \mathbb{P} \left( \widehat{Y}_j^{\pre} \neq Y_j \mid \widehat{Y}_j^{\pre} \in \calG_g \right) ,
\end{equation}
which represents the group-wise false classification probability in the first pre-classification step, and it can be viewed as the counterpart of the null proportion in the multiple testing procedures \citep{benjamini1995controlling,storey2004strong}.
Note that while the ``$+1$'' terms in \eqref{eq:proportion_RGg_est} lead to slight conservativeness compared to the direct proportion estimator $\sum_{ k\in \calG_g } | \widebar{\calR}_k | / \sum_{ k\in \calG_g } | \widebar{\calS}_k |$, this adjustment ensures the finite-sample error rate guarantee for large-scale inference, and it aligns with those in finite-sample multiple testing studies \citep{storey2004strong,barber2015controlling,lei2018adapt}.

\begin{remark}\label{remark:compare_conformal_infer}
The proposed PSP addresses the current classification problems through preliminary pre-classification and the conformalized uncertainty quantification for post-classification decisions.
Although large-scale testing methods with conformal $p$-values and BH procedure are widely studied in the literature \citep{mary2022semi,bates2023testing,
jin2023selection,gao2023simultaneous,marandon2024adaptive}, PSP introduces key distinctions in both the $p$-value construction and the large-scale inference procedure.
First, the $p$-values in \eqref{eq:pvalue} can constructed in a subtly selective way for the pre-classified subjects $\{ j\in[m]: j \in \calS_k, k \in \calG_g \}$ and scores $\{ \mu_k(X_{m+i}): i \in \widebar{\calR}_k, k \in \calG_g \}$.
In contrast to conventional conformal $p$-values that apply to all tested subjects and scores of hold-out data, the sets $\calS_k$ and $\widebar{\calR}_k$ in PSP are selected based on the pre-classification results.
Second, the procedure in \eqref{eq:threshold} is not simply the vanilla BH procedure, but is essentially a generalization for post-selection inference.
For each group $\calG_g,g\in[G]$, the subjects to be tested are randomly determined by the selective set $\bigcup_{ k\in\calG_g } \calS_k$.
It makes PSP apart form usual BH procedure where the tested subjects are typically deterministic.
\end{remark}

\section{Theoretical Results}\label{sec:theory}

In this section, we establish the theoretical results for the general group-wise PSP method proposed in Section \ref{sec:method}.
First, in Section \ref{sec:theory_error}, we present the finite-sample validity of $\FDR_{ \calG_g },g\in[G]$ control for Algorithm \ref{alg:method}.
Next, Section \ref{sec:theory_power} presents the power analysis for the proposed algorithm.
Specifically, we show that the post-classification decisions never degrade the power of pre-classification if it has been sufficiently accurate to meet the target error levels.  
Furthermore, we establish general optimality theories from both asymptotic and non-asymptotic perspectives for the proposed approach.

\subsection{Error Rate Control}\label{sec:theory_error}

This section establishes valid error rate control for the proposed PSP method in Algorithm \ref{alg:method}.
It is worthy noting that the result is distribution-free, with no specific restrictions or conditions imposed on the data distribution.

\begin{theorem}\label{thm:FDR}

Algorithm \ref{alg:method} achieves $\FDR_{ \calG_g } \leq \alpha_{ \calG_g }$ for $g\in [G]$.

\end{theorem}

Theorem \ref{thm:FDR} shows that the PSP method in Algorithm \ref{alg:method} achieves valid finite-sample $\FDR_{ \calG_g },g\in[G]$ guarantees at the target error levels.
In addition, the proposed PSP is model-free, allowing various machine learning classifiers to be flexibly integrated in the implementation of pre-classification rule $\texttt{PreClass}$ and score function $\mu$ in PSP.
Moreover, within the unified framework discussed in Section \ref{sec:problem}, Theorem \ref{thm:FDR} straightforwardly implies valid $\FDR_{ \tto }$ and $\FDR_k,k\in[K]$ control for the overall PSP in Algorithm \ref{alg:method_overall} and class-wise PSP in Algorithm \ref{alg:method_classwise}, respectively. 
The results are summarized in the following corollary.

\begin{corollary} 

Algorithm \ref{alg:method_overall} achieves $\FDR_{\tto} \leq \alpha$ for overall classification, and Algorithm \ref{alg:method_classwise} achieves $\FDR_k \leq \alpha_k, k\in [K]$ for class-wise classification.

\end{corollary}

\subsection{Power Analysis}\label{sec:theory_power}

\subsubsection{Power Non-Degradation}

Building on the error rate control results established in Section \ref{sec:theory_error}, we now focus on evaluating the power of the proposed PSP method.
Recall that PSP achieves error control by integrating the steps of pre-classification and post-classification decisions.
A potential concern with PSP at the first glance is that the cost of valid error control might be a power loss of the pre-classification.
However, as demonstrated in Theorem \ref{thm:power_nondegrade}, the step of post-classification decisions facilitates valid error rate guarantees without compromising the power of the pre-classification.

\begin{theorem}\label{thm:power_nondegrade}

For each $g\in[G]$ and group $\calG_g$, Algorithm \ref{alg:method} yields $\widehat{Y}_j = \widehat{Y}_j^{\pre}$ for all $j\in \bigcup_{ k\in\calG_g } \calS_{k}$ if the empirical group-wise false pre-classification rate satisfies $\widehat{\theta}_{ \calR, \calG_g } \leq \alpha_{ \calG_g }$.

\end{theorem}

Theorem \ref{thm:power_nondegrade} shows that the post-classification decisions in Algorithm \ref{alg:method} will never degrade the accuracy of the pre-classification step, provided that the pre-classification has been sufficiently powerful to directly meet the target error levels.
This non-degradation property ensures the safe employment of PSP in practice, as indecisions will only be made to avoid the high misclassification in the cases of ambiguity or low confidence when the accuracy of pre-classification rule $\texttt{PreClass}$ does not meet the target error level.

Next, to extend the power analysis beyond the cases where a sufficiently powerful pre-classification procedure is available, we present the general power analysis for the proposed PSP approach in the subsequent section.

\subsubsection{Power Optimality}\label{sec:theory_power_optimality}

In this section, we establish the general theories for the power optimality of the PSP approach in Algorithm \ref{alg:method} from both non-asymptotic and asymptotic perspectives.
To this end, we first aim to derive the oracle optimal procedure that maximizes the power while maintaining error rate control, and then study the approximation of the proposed PSP to the oracle.
Due to the fact that the pre-classification can possibly be predetermined or derived from an external classifier in practice, we freeze $\texttt{PreClass}$ in the pre-classification step and focus on optimizing the post-classification decisions.
Therefore, for group $\calG_g$, given the pre-classification results $\widehat{Y}_j^{\pre},j \in \bigcup_{ k \in \calG_g } \calS_k$, the goal of the optimal procedure is to optimize the decisions $(\delta_j)_{j\in \bigcup_{ k \in \calG_g } \calS_k} \in \left\{ 0, 1 \right\}^{\sum_{k\in\calG_g} |\calS_k|}$ such that the post-classification results $\widehat{Y}_j = \widehat{Y}_j^{\pre} \delta_j, j\in \bigcup_{ k \in \calG_g } \calS_k$ maximize the number of true classifications while ensuring the group-wise error rate guarantee.

To facilitate the optimality analysis, we define the group-wise marginal false decision rate (mFDR) as surrogate to the $\FDR_{ \calG_g }$ in \eqref{eq:FDR_g}:
$$
\mFDR_{ \calG_g } = \frac{ \bbE \sum_{j\in[m]} \bbI\left( \widehat{Y}_j \in \calG_g, \widehat{Y}_j \neq Y_j \right) }{ \bbE \sum_{j\in[m]} \bbI\left( \widehat{Y}_j \in \calG_g \right)} , ~ g \in [G] ,
$$
and similar surrogates are commonly adopted in multiple testing problems \citep{genovese2002operating,sun2007oracle,
lei2018adapt,cao2022optimal} for theoretical analysis.
Then, the optimal procedure for group $\calG_g$ can be formulated as the following optimization:
\begin{equation}\label{eq:power}
\begin{aligned}
\max_{ (\delta_j)_{ j\in \bigcup_{ k \in \calG_g } \calS_k } }  & \bbE \sum_{ j\in\calQ_{\calG_g} }  \bbI\left( \delta_j = 1 \right) \\
\text{s.t. }  & \mFDR_{ \calG_g } = \frac{ \bbE \sum_{ k\in\calG_g } \sum_{ j\in \calS_k } \bbI\left( \delta_j = 1, Y_j \neq k \right) }{ \bbE \sum_{ k\in\calG_g } \sum_{ j\in \calS_k } \bbI\left( \delta_j = 1 \right)} \leq \alpha_{ \calG_g } ,
\end{aligned}
\end{equation}
where $\delta_j = \delta_j(X_j,\widehat{Y}_j^{\pre}): \calX \times [K] \rightarrow \{0,1\}$ represents the decision rule for $j\in \bigcup_{ k \in \calG_g } \calS_k$, and 
$\calQ_{ \calG_g } = \bigcup_{ k \in \calG_g } \{ j \in \calS_k : Y_j = k \}$ denotes the index set of the true pre-classification for group $\calG_g$.
The optimization procedure \eqref{eq:power} serves as the optimal baseline for the power analysis.

Before presenting the optimality theories for the proposed Algorithm \ref{alg:method}, several definitions are introduced. 
Define the posterior probabilities $\mu^{\star}:= (\mu_1^{\star}, \cdots, \mu_K^{\star})$, where
\begin{equation}\label{eq:mu_star}
\mu_k^{\star}(x) = \mathbb{P} \left( Y_j = k \mid X_j = x \right), 
\end{equation}
and define the corresponding group-wise probability distribution functions
\begin{equation}\label{eq:F0}
F_{0, \calG_g}(t) = \mathbb{P}\left( \mu_{ \widehat{Y}_j^{\pre} }^{\star}(X_j) > t \mid \widehat{Y}_j^{\pre} \in \calG_g, \widehat{Y}_j^{\pre} \neq Y_j \right), ~ t\in\bbR,
\end{equation}
\begin{equation}\label{eq:F1}
F_{1, \calG_g}(t) = \mathbb{P}\left( \mu_{ \widehat{Y}_j^{\pre} }^{\star}(X_j) > t \mid \widehat{Y}_j^{\pre} \in \calG_g \right), ~ t\in\bbR .
\end{equation}
Then the following theorem derives the optimal decision rules for optimization \eqref{eq:power}.

\begin{theorem}\label{thm:power_opt_oracle}

For group $\calG_g$, define $R_{ \calG_g }(t) = \frac{ \theta_{ \calR, \calG_g } F_{0, \calG_g}(t) }{ F_{1, \calG_g}(t) }$, where $\theta_{ \calR, \calG_g }$ is given in \eqref{eq:proportion_RGg}. 
Assume that there exist $\widebar{t}_{ \calG_g } \in [0,1)$ satisfying $R_{ \calG_g }( \widebar{t}_{ \calG_g } ) \leq \alpha_{ \calG_g }$ such that $t_{\calG_g}^{\star} := \inf \left\{ t \geq 0: R_{ \calG_g } (t) \leq \alpha_{ \calG_g } \right\}$ is well-defined, and assume that $F_{0, \calG_g}$ and $F_{1, \calG_g}$ are continuous at $t = 0$ and $t = t_{\calG_g}^{\star}$.
Then, the optimal solution to procedure \eqref{eq:power} is given by $\delta_j^{\opt} = \bbI ( \mu_{ \widehat{Y}_j^{\pre} }^{\star}(X_j) > t_{ \calG_g }^{\star} )$ for $j\in \bigcup_{ k \in \calG_g } \calS_k$.
\end{theorem}

Theorem \ref{thm:power_opt_oracle} shows that the optimal group-wise decisions are the threholding rules based on the functions $\{ \mu_{ k }^{\star}(\cdot) \}_{ k\in\calG_g }$ in \eqref{eq:mu_star} and threshold $t_{\calG_g}^{\star}$ defined in Theorem \ref{thm:power_opt_oracle}.
While similar posterior thresholding rules have been derived in model-free inference problems to guide the selection of score functions in practice \citep{lei2018adapt,zhao2025false}, 
however, the optimality of the inference procedures is not guaranteed solely by the choice of score function as potential sub-optimality may arise from the data-driven determination of the threshold.
Consequently, we proceed to establish a comprehensive analysis for the approximation of PSP to the oracle optimal rules.

For the score function $\mu = (\mu_1, \cdots, \mu_K)$ employed in the PSP Algorithm, define 
\begin{equation}\label{eq:epsilon}
\epsilon_{\calG_g} = \min_{ h \in \calH } \max_{k\in\calG_g, j\in[m+n]} \left| \mu_k^{\star}(X_j) - h \left( \mu_k(X_j) \right) \right|
\end{equation}
to be the group-wise deviation error between $\mu$ and the oracle $\mu^{\star}$ up to a monotone transformation, where $\calH$ represents the set of all strictly increasing functions $h: \bbR \rightarrow \bbR$, and define 
\begin{equation}\label{eq:D_t_eps}
D_{a, \calG_g}(t,\epsilon_{\calG_g}) = \sup_{ t^\prime \in[0, t] } \left( F_{a, \calG_g}(t^\prime-\epsilon_{\calG_g}) - F_{a, \calG_g}(t^\prime+\epsilon_{\calG_g}) \right) , ~ t \in (0,1) , ~ a\in\{0,1\}
\end{equation}
to be the perturbations of the probability functions $F_{0, \calG_g}$ and $F_{1, \calG_g}$ in \eqref{eq:F0}-\eqref{eq:F1} induced by the deviation $\epsilon_{\calG_g}$.
Notably, the definition of $\epsilon_{\calG_g}$ allows for deviation up to a monotone transformation, since the constructed $p$-values in \eqref{eq:pvalue} are invariant under any strictly increasing transformation $h \in \calH$.
The following Theorem \ref{thm:power_opt_approx} establishes the optimality of the proposed PSP by characterizing its approximation to the oracle from Theorem \ref{thm:power_opt_oracle}.

\begin{theorem}\label{thm:power_opt_approx}

Suppose that the assumptions in Theorem \ref{thm:power_opt_oracle} hold. 

\begin{enumerate}[nosep, label=\thetheorem.\alph*]

\item \label{thm:power_opt_approx_nonasymp}
Non-Asymptotic Approximation to the Oracle:
Define $\theta_{ \calS, \calG_g } = \bbP ( \widehat{Y}_{j}^{\pre} \in \calG_g )$, and let $\widetilde{T}_{ \calG_g }$ be the threshold of procedure \eqref{eq:threshold} with error level $\widetilde{\alpha}_{ \calG_g }$.
Suppose that $D_{1, \calG_g}(1 - \alpha_{ \calG_g }, \epsilon_{\calG_g}) \leq \frac{1}{4} F_{1, \calG_g}(1 - \alpha_{ \calG_g })$, then for any $\Delta > 0$ satisfying $\Delta \leq \frac{1}{2} \theta_{ \calS, \calG_g } \sqrt{ \min(m,n) }$, $\Delta \leq \frac{1}{2\sqrt{2}} \theta_{ \calR, \calG_g } \sqrt{ \theta_{ \calS, \calG_g } n }$ and $\Delta \leq \frac{1}{4\sqrt{2}} F_{1, \calG_g}( 1 - \alpha_{ \calG_g } ) \sqrt{ \theta_{ \calS, \calG_g } m }$, we have
$$
\bbI\left( p_j \leq \widetilde{T}_{ \calG_g } \right) \geq \delta_j^{\opt}, ~ j\in \bigcup_{ k\in\calG_g } \calS_k,
$$
holds with probability at least $1 - 10 \exp\left( - 2\Delta^2 \right)$, where
\begin{equation}\label{eq:alpha_inflate}
\widetilde{\alpha}_{ \calG_g } = \alpha_{ \calG_g } \left( 1 +  \sup_{t\in[0,1-\alpha_{ \calG_g }]} \frac{ F_{0, \calG_g}(t - \epsilon_{ \calG_g }) - F_{0, \calG_g}(t) }{F_{0, \calG_g}(t) } \right) + \eta_{ \calG_g } ,
\end{equation}
and
\begin{align*}
\eta_{ \calG_g } = & \frac{ 1 }{  F_{1, \calG_g}( 1-\alpha_{ \calG_g } ) } \left( \frac{ ( 6 \sqrt{ \theta_{ \calR, \calG_g } } + \sqrt{2} ) \Delta }{ \sqrt{  \theta_{ \calS, \calG_g } n } } + \frac{ 3\sqrt{2} \Delta }{ \sqrt{ \theta_{ \calS, \calG_g } m } } + \frac{ 12 + 2\theta_{ \calR, \calG_g } }{ \theta_{ \calS, \calG_g } n } \right. \\
& \left. + 3 \theta_{ \calR, \calG_g }  D_{0,\calG_g}(1-\alpha_{ \calG_g }, \epsilon_{\calG_g}) +  3 D_{1,\calG_g}(1-\alpha_{ \calG_g }, \epsilon_{\calG_g}) \right) .
\end{align*}
Consequently, $\sum_{j \in \calQ_{ \calG_g } } \bbI ( p_j \leq \widetilde{T}_{ \calG_g } ) \geq \sum_{j \in \calQ_{ \calG_g } } \delta_j^{\opt}$ holds with high probability.

\item \label{thm:power_opt_approx_asymp}
Asymptotic Approximation to the Oracle:
Suppose that there exists $\xi_0>0$ such that $R_{ \calG_g }(t) < \alpha_{ \calG_g }$ for all $t \in (t_{ \calG_g }^{\star}, t_{ \calG_g }^{\star} + \xi_0 )$ if $R_{ \calG_g }(t_{ \calG_g }^{\star}) = \alpha_{ \calG_g }$.
For fixed level $\alpha_{ \calG_g } \in(0,1)$ and fixed distribution of $(X_j,Y_j,U_j),j\in[m+n]$, if $D_{0,\calG_g}(1-\alpha_{ \calG_g }, \epsilon_{\calG_g}) = o_{\pr} (1)$ and $D_{1,\calG_g}(1-\alpha_{ \calG_g }, \epsilon_{\calG_g}) = o_{\pr} (1)$ as $(m,n) \rightarrow \infty$, then 
$$
\frac{1}{|\calQ_{ \calG_g }|} \sum_{ j\in \calQ_{\calG_g} } \bbI \left( p_j \leq \widehat{T}_{ \calG_g } \right) = \frac{1}{|\calQ_{ \calG_g }|} \sum_{ j \in \calQ_{\calG_g} } \delta_j^{\opt} + o_{\pr} (1) .
$$
\end{enumerate}
\end{theorem}

The non-asymptotic analysis in Theorem \ref{thm:power_opt_approx_nonasymp} shows that if applied with a slightly inflated error level $\widetilde{\alpha}_{\calG_g}$ in \eqref{eq:alpha_inflate}, the PSP method in Algorithm \ref{alg:method} achieves greater power than the oracle optimal rules with high probability.
The difference between the inflated level $\widetilde{\alpha}_{\calG_g}$ and target level $\alpha_{ \calG_g }$ mainly depends on the sample sizes $(m,n)$ of target data and hold-out data and the deviation $\epsilon_{\calG_g}$ in \eqref{eq:epsilon}.
The inflation effect diminishes as the sample sizes increase.
Moreover, the impact of deviation $\epsilon_{\calG_g}$ is negligible if the oracle score function is accurately approximated and the distribution functions $F_{0, \calG_g}$ and $F_{1, \calG_g}$ are continuous over $[0, 1-\alpha_{\calG_g}]$.
We note that the constant terms in Theorem \ref{thm:power_opt_approx_nonasymp} are not optimized, as the current result is sufficient for understanding the non-asymptotic optimality of the proposed PSP approach.

In addition, the asymptotic analysis in Theorem \ref{thm:power_opt_approx_asymp} further establishes the large-sample asymptotic convergence of the power of Algorithm \ref{alg:method} to the oracle, provided that the perturbations $D_{0,\calG_g}(1-\alpha_{ \calG_g }, \epsilon_{\calG_g})$ and $D_{1,\calG_g}(1-\alpha_{ \calG_g }, \epsilon_{\calG_g})$ in \eqref{eq:D_t_eps} vanish in asymptotically.
The conditions essentially require the continuity of functions $F_{0, \calG_g}$ and $F_{1, \calG_g}$ and that $\mu$ converges to the oracle posterior $\mu^{\star}$ up to a monotone transformation, which aligns with the implications for the non-asymptotic result in Theorem \ref{thm:power_opt_approx_nonasymp}.

In summary, Theorems \ref{thm:power_opt_oracle}-\ref{thm:power_opt_approx} demonstrate the optimality of the proposed PSP method for the group-wise classification from both non-asymptotic and asymptotic perspectives. 
In addition to guiding the choice of score function by approximating the oracle scores as in the model-free inference problems \citep{lei2018adapt,zhao2025false}, we further provide a rigorous and comprehensive analysis of the approximation between the derived decision rules and the oracle optimality.
This contributes to advancing the theoretical analysis and understanding of model-free inference.
Moreover, due to the incorporation of conformalization in constructing the selective $p$-value \eqref{eq:pvalue}, the derived optimality results shed light on the power of conformal inference approaches.

\begin{remark}
The deviation $\epsilon_{\calG_g}$ of the score function represents a structure-agnostic approximation error \citep{balakrishnan2023fundamental}.
Unlike conventional convergence rates in statistical learning that depend on specific structural regularities of target function, the structure-agnostic analysis circumvents such assumptions.
Moreover, estimation convergence is typically tied to training sample sizes, while modern machine learning algorithms perform well through fine-tuned pre-trained models on small task-specific data or few-shot generalization \citep{brown2020language}.
Thus, $\epsilon_{\calG_g}$ characterizes the general model-free approximation.
\end{remark}

\section{Simulation Study}\label{sec:simulation}

In this section, we validate the proposed PSP Algorithm \ref{alg:method} through simulation study, focusing on the overall and class-wise classifications as introduced in Section \ref{sec:problem}.
We generate i.i.d. target data $((X_j,Y_j))_{j\in[m]}$ and hold-out data $((X_{m+i},Y_{m+i}))_{i\in[n]}$ from $d$-dimensional Gaussian mixture models. 
The class probabilities $P(Y_j = k),k\in[K]$ are given by $( Z_1,\cdots,Z_K ) / \sum_{k\in[K]} Z_k$, where $Z_k \overset{\text{i.i.d.}}{\sim} \text{Uniform}(1,2)$, and the conditional distributions follow $X_j \mid Y_j = k \sim \text{N}(m_k, I_d)$, where $I_d$ is the identity matrix and $m_k\in\bbR^d$ represents the mean vectors for class $k\in[K]$.
Additionally, we generate a training dataset of size $n_{\text{tr}}$ to train \texttt{PreClass} for pre-classification and to train score function $\mu$ for constructing selective $p$-values.
Throughout the simulations, we set $m=n=n_{\text{tr}} = K n_0$ with $n_0 = 100$, $d = 10$, and $m_k = k \times 1_d / d^{1/4}$ for $k\in[K]$. 
Each simulation consists of 500 independent replications to evaluate the results.

\begin{figure}[t]
\centering
\includegraphics[scale=0.5]{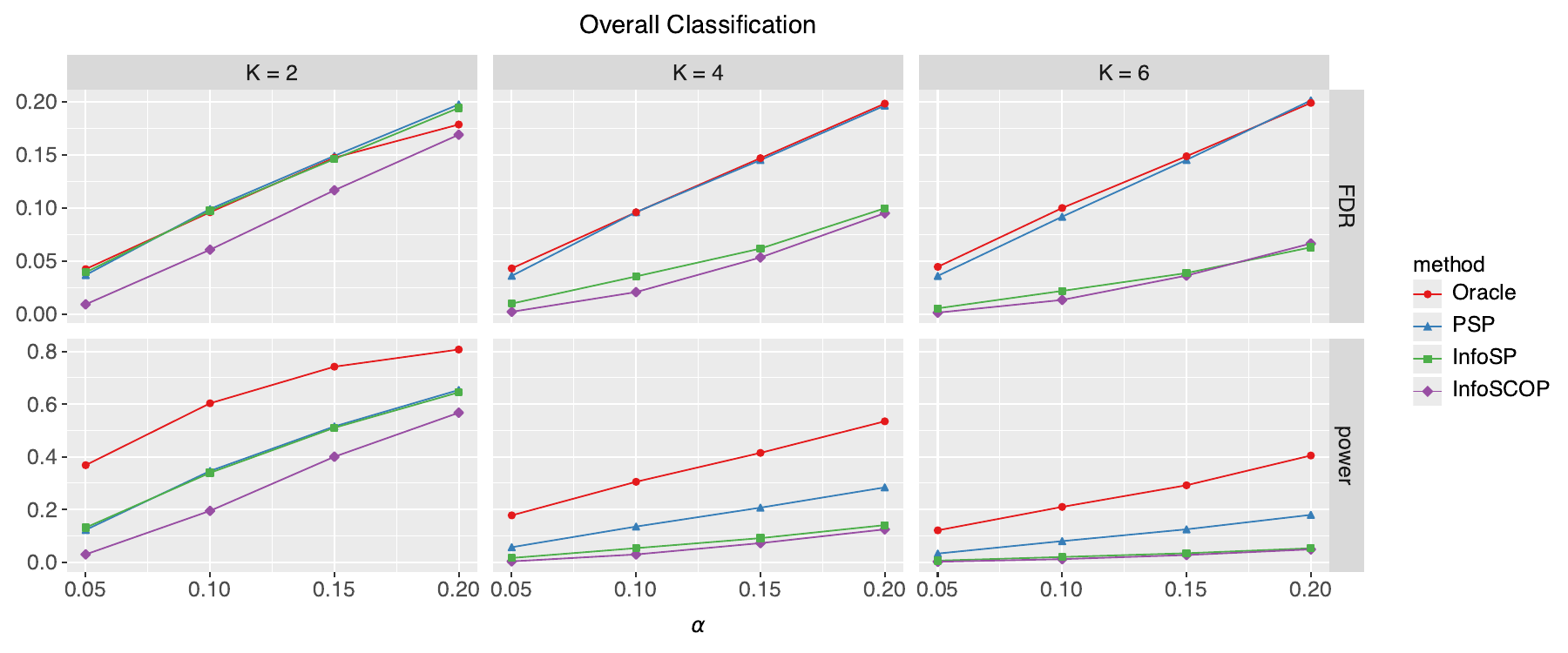}
\caption{Empirical FDR and power comparisons for \texttt{Oracle}, \texttt{PSP}, \texttt{InfoSP} and \texttt{InfoSCOP} in overall classification problems. 
}
\label{fig:simu_overall_lgbm}
\end{figure}

\begin{figure}[t]
\centering
\includegraphics[scale=0.5]{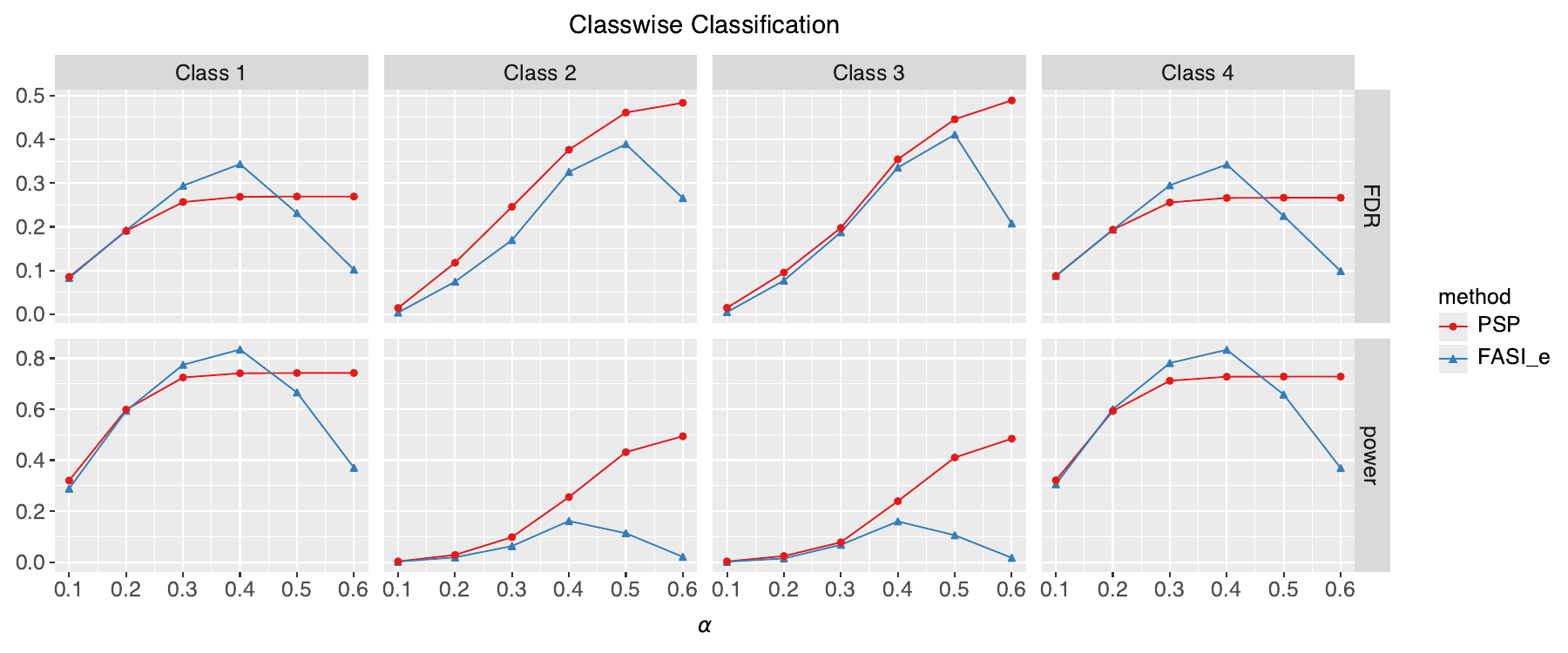}
\caption{Empirical classwise FDR and power comparisons for \texttt{PSP} and \texttt{FASI\_e} in class-wise classification problems. 
}
\label{fig:simu_classwise_lgbm}
\end{figure}

We first evaluate the performance of overall PSP in Algorithm \ref{alg:method_overall}. 
For classification with overall misclassification, \citet{cho1970optimum,ndaoud2024ask} show that the oracle decision rule is derived by thresholding the maximal posterior class probability.
As practical methods for estimating the oracle rule are lacking therein, we incorporate this rule into PSP to set $\mu^{\pre} = \mu = \mu^{\star}$ and $\text{PreClass}(x) = \arg\max_{k \in [K]} \mu_k^{\star}(x)$ as oracle baseline, denoted as \texttt{Oracle}, for comparison.
We also implement PSP using learned score functions, denoted as \texttt{PSP}.
Additionally, \citet{gazin2025selecting} studies informative conformal prediction set selection and construction, and introduces \texttt{InfoSP} and \texttt{InfoSCOP} approaches with overall error rate control.
We also implement both \texttt{InfoSP} and \texttt{InfoSCOP} by defining the informative prediction sets as those containing only one class label. 
We set $K \in \{ 2,4,6 \}$, and score functions in \texttt{PSP}, \texttt{InfoSP} and \texttt{InfoSCOP} are learned from the training dataset by LightGBM classifier, expect for \texttt{Oracle} where the oracle $\mu^{\star}$ is employed.
Figure \ref{fig:simu_overall_lgbm} presents the results on overall FDRs and powers across the target error levels $\alpha \in \{ 0.05, 0.1, 0.15, 0.2 \}$, where the power is assessed by $\sum_{ j\in [m] } \bbI ( \widehat{Y}_j = Y_j ) / m$.
The results obtained by random forest and SVM score training are similar and are provided in Section \ref{sec:add_simu} of the supplement for brevity.
Figure \ref{fig:simu_overall_lgbm} shows that \texttt{Oracle} achieves the highest power.
When $K =2$, \texttt{PSP} and \texttt{InfoSP} perform similarly, and \texttt{InfoSCOP} shows inferior performance that is consistent with findings in \citet{gazin2025selecting}.
However, when $K >2$, both \texttt{InfoSP} and \texttt{InfoSCOP} show reduced power.
These results demonstrate the power advantage of PSP in overall classifications.

Next, we evaluate the class-wise PSP in Algorithm \ref{alg:method_classwise}.
For PSP, we implement the same pre-classification as in the overall classification case and denote it as \texttt{PSP}.
As reviewed in Section \ref{sec:intro_review}, \citet{rava2024burden} develops a conformal approach, named FASI, for binary classification with error control for one specific class.
We remark that although it can be extended to multi-class data for a single target class, it does not support simultaneous classification across multiple classes.
Besides, it remains unclear how the updated FASI in \citet{rava2025burden} can be extended to multi-class classification, as also noted in Section G of \citet{rava2025burden}.
To enable comparison, we implement FASI from \citet{rava2024burden} for each class separately and make the final classification only for subjects assigned to exactly one class, denoting this extension as \texttt{FASI\_e}.
In other words, if a subject is assigned to multiple classes or left with indecisions for all classes, \texttt{FASI\_e} ultimately makes an indecision.
We set $K = 4$, train scores in \texttt{PSP} and \texttt{FASI\_e} by LightGBM, and measure the class-wise power by $\sum_{ j \in [m] } \bbI ( \widehat{Y}_j = Y_j = k ) / \sum_{j\in[m]} \bbI \left( Y_j = k \right)$ for $k\in[K]$.
Figure \ref{fig:simu_classwise_lgbm} presents the class-wise FDRs and powers for \texttt{PSP} and \texttt{FASI\_e} across target error levels $\alpha_1 = \cdots = \alpha_K$ from 0.1 to 0.6. 
Similar results with scores learned by random forest and SVM classifiers are presented in Section \ref{sec:add_simu} of the supplement.
It is shown that \texttt{FASI\_e} outperforms \texttt{PSP} at moderate error levels for classes 1 and 4, but suffers substantial power loss as the error level increases. 
This is contrary to the general knowledge that power increases with higher error levels, as seen in \texttt{PSP}. 
The reason is that FASI in \citet{rava2024burden} is designed for single-class error control.
As the target error level increases, it tends to assign multiple class labels to a subject, where \texttt{FASI\_e} will make an indecision.
Therefore, PSP serves as a feasible classification algorithm with general error rate control.

\section{Real Data Analysis}\label{sec:realdata}

In this section, we apply the proposed PSP approach to CIFAR-10 dataset to evaluate its performance.
CIFAR-10 is a widely used benchmark dataset in computer vision, containing 50,000 training images and 10,000 test images across $K = 10$ classes.
We first split the training images into two subsets: 70\% for training a base classifier and 30\% as the hold-out set.
Based on the trained base classifier, we then implement the overall and class-wise PSP approaches in Algorithms \ref{alg:method_overall}-\ref{alg:method_classwise} for overall and class-wise classifications, respectively.

\begin{figure}[t]
\centering
\includegraphics[scale=0.5]{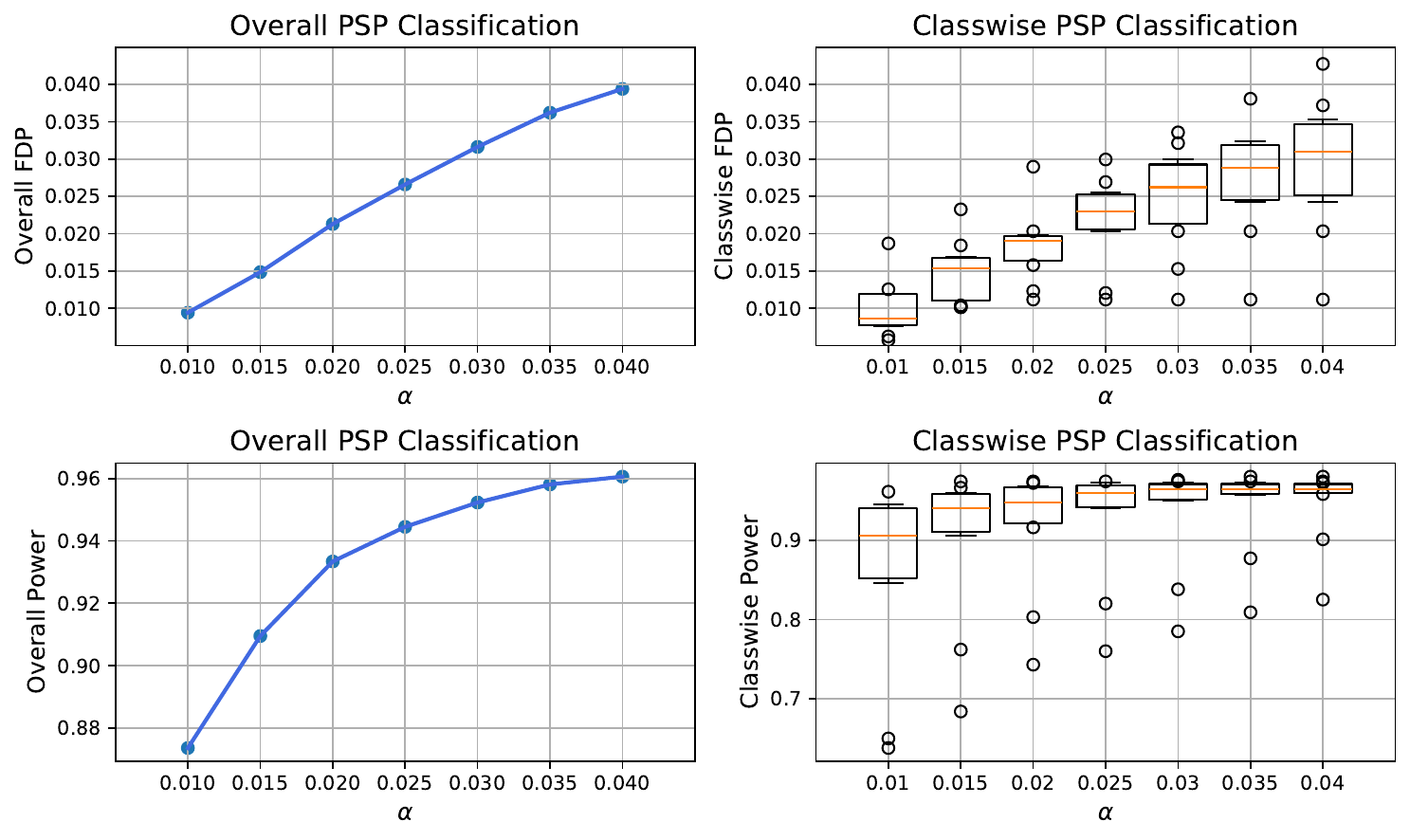}
\caption{Results of PSP methods on CIFAR-10. The top-left and bottom-left panels show the overall $\FDP$ and power for the overall PSP approach.
The top-right and bottom-right panels show the boxplots of class-wise $\FDP_k,k\in[K]$ and powers for the class-wise PSP approach. }
\label{fig:realdata}
\end{figure}

First, by using 70\% of the CIFAR-10 training images, we begin by fine-tuning a Vision Transformer (ViT)-base model \citep{dosovitskiy2020image} that is pre-trained on the external ImageNet-21k dataset.
We freeze the pre-trained ViT except for the final classifier layer that will be updated in the fine-tuning.
The obtained model that outputs probabilities for each of the 10 classes is employed for both $\mu^{\pre}$ in the pre-classification step and $\mu$ in the selective $p$-value construction step.
The pre-classification rule \texttt{PreClass} assigns each image in hold-out and testing datasets to the class with highest output probability.
As a result, it achieves an overall false pre-classification rate of 3.9\% on the testing dataset, and the class-wise false pre-classification rates range from 1.1\% to 10.6\%.
Next, we leverage the 30\% hold-out data to implement the overall PSP in Algorithm \ref{alg:method_overall} with error level $\alpha$ and the class-wise PSP in Algorithm \ref{alg:method_classwise} with error levels $\{ \alpha_k \}_{k\in[K]}$, separately.
In our experiments, we set $\alpha$ and $\alpha_1 = \cdots = \alpha_K$ ranging from 0.01 to 0.04, and the results of PSP on testing images are presented in Figure \ref{fig:realdata}.
The top-left panel shows the $\FDP$ for overall classification, and the top-right panel shows the boxplots of $\FDP_k,k\in[K]$ for class-wise classification.
Besides, the bottom-left and bottom-right panels show the corresponding overall and class-wise powers, respectively.
These results validate the effectiveness of the proposed algorithm in ensuring valid error rate guarantees while maintaining classification power.

\section{Inference with $e$-Values}\label{sec:evalue}

Preceding sections introduce the PSP algorithm and present the theoretical and numerical results for PSP.
Notably, large-scale inference approaches are often associated with $e$-value-based inference \citep{wang2022false,ren2024derandomised,bashari2023derandomized,li2025note}. 
In this section, we develop an ePSP algorithm that integrates the idea of PSP with selective $e$-values, which enables prospective applications in aggregated inference.

Similar to the methodology in Section \ref{sec:method}, PSP with $e$-values (ePSP) also consists of three steps: pre-classification, selective $e$-value construction, and large-scale post-classification decisions. 
The detailed algorithm is as follows:
\begin{enumerate}[nosep]
\item Pre-classification. Implement the step of pre-classification in Section \ref{sec:method_preclass} to obtain the pre-classification labels $\{ \widehat{Y}_j^{\pre} \}_{ j\in[m+n] }$ and sets $S_k$ and $\widebar{S}_k$ for $k\in[K]$.

\item Selective $e$-value construction. For $g\in[G]$ and group $\calG_g$, construct the selective $e$-values as
\begin{equation}\label{eq:evalue}
e_j = \frac{ 1 + \sum_{ k^\prime \in \calG_g } | \widebar{\calR}_{ k^\prime } |  }{ \widehat{\theta}_{ \calR, \calG_g } } \frac{ \bbI ( \mu_k( X_j ) \geq \widehat{t}_{ \calG_g }^{\e} ) }{ 1 
+ \sum_{ k^\prime \in \calG_g } \sum_{ i \in \widebar{\calR}_{k^\prime} } \bbI ( \mu_{k^\prime} (X_{m+i}) \geq \widehat{t}_{ \calG_g }^{\e} ) } , ~ j \in \calS_k , ~ k\in \calG_g ,
\end{equation}
where the threshold $\widehat{t}_{ \calG_g }^{\e}$ is determined by
\begin{equation}\label{eq:threshold_e}
\widehat{t}_{ \calG_g }^{\e} = \min \left\{ t \in \calT_{ \calG_g }: \frac{ \sum_{ k\in \calG_g } |\calS_k| }{ 1 + \sum_{ k \in \calG_g } | \widebar{\calR}_k | } \frac{ 1 
+ \sum_{ k \in \calG_g } \sum_{ i \in \widebar{\calR}_k } \bbI ( \mu_k (X_{m+i}) \geq t ) }{ \sum_{ k\in \calG_g } \sum_{ j \in \calS_k } \bbI ( \mu_k( X_j ) \geq t ) } \leq \frac{ \alpha_{ \calG_g }^{\prime} }{ \widehat{\theta}_{ \calR,  \calG_g } } \right\} 
\end{equation}
with
\begin{equation}\label{eq:Tt_set}
\calT_{ \calG_g } =  \{  \mu_k( X_j ) \}_{ j \in \calS_k, k\in \calG_g} \bigcup \{ \mu_{k} (X_{m+i}) \}_{ i \in \widebar{\calR}_k, k\in \calG_g } ,
\end{equation}
and $\alpha_{ \calG_g }^{\prime} \in (0,1)$ is a pre-determined level.
Set $\widehat{t}_{ \calG_g }^{\e} = +\infty$ if the $\widehat{t}_{ \calG_g }^{\e}$ in \eqref{eq:threshold_e} does not exist.

\item Large-scale post-classification decisions. For each $g\in[G]$ and group $\calG_g$, apply the eBH procedure \citep{wang2022false} with the selective $e$-values $\{e_j\}_{ j \in \bigcup_{ k\in \calG_g } \calS_k }$ at the target level $\alpha_{ \calG_g }$: 
\begin{equation}\label{eq:threshold_e_l}
\widehat{T}_{\calG_g}^{\e} = e_{( \calG_g, \widehat{l}_{\calG_g}^{\e} )} , \text{ with } \widehat{l}_{\calG_g}^{\e} = \max\left\{ l\in [ \sum_{k\in \calG_g} |\calS_k| ] : e_{(\calG_g,l)} \geq  \sum_{k\in \calG_g} |\calS_k| / (l \alpha_{ \calG_g }) \right\} ,
\end{equation}
where $e_{(\calG_g,1)} \geq \cdots \geq e_{(\calG_g,\sum_{k\in \calG_g} |\calS_k|)}$ denote the sorted values of $(e_j)_{ j \in \bigcup_{ k\in \calG_g } \calS_k }$ in descending order, and set $\widehat{T}_{\calG_g}^{\e} = +\infty$ if the $\widehat{l}_{\calG_g}^{\e}$ above does not exist.
Let 
\begin{equation}\label{eq:postclass_decision_e}
\widehat{Y}_j^{\e} = \widehat{Y}_j^{\pre} \bbI \left(  e_j \geq \widehat{T}_{\calG_g}^{\e} \right), ~ j\in \bigcup_{ k\in\calG_g } \calS_k .
\end{equation}
\end{enumerate}

The following Theorem \ref{thm:FDR_e} presents the properties of ePSP algorithm.

\begin{theorem}\label{thm:FDR_e}
For each $g\in[G]$:
\begin{enumerate}[nosep, label=\thetheorem.\alph*]
\item \label{thm:FDR_e_control}
The $\FDR_{ \calG_g }$ induced by $( \widehat{Y}_j^{\e} )_{ j\in[m] }$ in ePSP satisfies $\FDR_{ \calG_g } \leq \alpha_{ \calG_g }$.

\item \label{thm:FDR_e_equiv}
If $\alpha_{ \calG_g }^{\prime} = \alpha_{ \calG_g }$, ePSP yields the same results as PSP in Algorithm \ref{alg:method}, i.e., $\widehat{Y}_j^{\e} = \widehat{Y}_j$ for all $\bigcup_{ k\in\calG_g } \calS_k$.

\item \label{thm:FDR_e_inferior}
For any $\alpha_{ \calG_g }^{\prime}$, $\sum_{ j\in\bigcup_{ k\in\calG_g } \calS_k }\bbI ( \widehat{Y}_j^{\e} \neq 0 ) \leq \sum_{ j\in\bigcup_{ k\in\calG_g } \calS_k }\bbI ( \widehat{Y}_j \neq 0 )$.
\end{enumerate}

\end{theorem}

Theorem \ref{thm:FDR_e_control} establishes the valid error rate control guaranteed by ePSP.
Furthermore, Theorems \ref{thm:FDR_e_equiv}-\ref{thm:FDR_e_inferior} presents the connections between ePSP and PSP, and implies that the admissible choice of $\alpha_{ \calG_g }^{\prime}$ in ePSP is just $\alpha_{ \calG_g }^{\prime} = \alpha_{ \calG_g }$, leading to identical decisions as PSP.
Similar inferiority is also observed in Knockoff inference \citep{ren2024derandomised}, whereas the employment of $e$-values facilitates the aggregated inference through multiple $e$-value averaging \citep{ren2024derandomised,bashari2023derandomized}.
The selective $e$-values constructed in ePSP can also be incorporated into aggregation procedures for inference, and related algorithm is presented in Section \ref{sec:ePSP} of the supplement for brevity.

\section{Extensions}\label{sec:extend}

This article introduces the inference framework of PSP for classifications with error rate control.
Beyond its primary application, PSP also has the potential for broader applications.
In this section, we discuss how to extend and integrate the idea of PSP into other problems for large-scale inference.
In the following, we focus on the problems of subject selection \citep{jin2023selection} and the construction of informative prediction sets \citep{gazin2025selecting} for illustration.

\subsection{Subject Selection}\label{sec:extend_subject}

Given the unlabelled covariates $X_1,\cdots,X_m \in \calX$ of interest and labelled data $(X_{m+i},Y_{m+i}) \in \calX \times \calY, i\in[n]$, the task of subject selection aims to select a subset of $[m]$ such that the corresponding unobserved labels satisfy a certain desirable criterion \citep{jin2023selection}.
\citet{jin2023selection} formulates the problem as large-scale hypothesis testing and develops the conformal selection approach to address it.
Next we show how PSP can be employed to develop flexible and powerful procedures for subject selection.

Denote by $\calC \subset \calY$ the region of interest in $\calY$, and the problem can be formulated as the hypotheses ``$\calH_{0,j}: Y_j \notin \calC$ versus $\calH_{1,j}: Y_j \in \calC$'' for $j\in[m]$, and the goal is to select a subset $\calS \subset [m]$ such that $\bbE \sum_{ j\in\calS } \bbI \left( Y_j \notin \calC \right) / \left( 1\vee |\calS| \right) \leq \alpha$.
Let $\mu(x):\calX \rightarrow [0,1]$ be a pre-trained classifier that learns the probability $\bbP\left( Y_j \in \calC \mid X_j = x \right)$ as the evidence whether the response corresponding to the covariate lies in the region of interest, and let $\texttt{Pre}(x):\calX \rightarrow \{0,1\}$ be a pre-trained pre-selection rule.
Then PSP can be implemented in the following three steps: 
\begin{enumerate}[nosep]
\item Pre-selection. $\calS^{\pre} = \{ j\in[m]: \texttt{Pre}(X_j) = 1 \}$ and $\widebar{\calS}^{\pre} = \{ i\in[n]: \texttt{Pre}(X_{m+i}) = 1 \}$.

\item Selective $p$-value construction. $p_j = \frac{ 1 }{ 1+|\widebar{\calR}| } \left( 1 + \sum_{ i\in \widebar{\calR} } \bbI \left( \mu(X_{m+i}) \geq \mu(X_{j}) \right) \right)$ for $j\in\calS^{\pre}$, where $\widebar{\calR} = \{ i\in \widebar{\calS}^{\pre} : Y_{m+i} \notin \calC \}$.

\item Post-selection inference. $\calS = \{ j\in\calS^{\pre}: p_j \leq p_{(\widehat{l})} \}$, where 
$
\widehat{l} = \max \{ l \in [|\calS^{\pre}|]: p_{(l)} \leq \frac{ l \alpha }{ \widehat{\theta} |\calS^{\pre}| } \}
$
and $\widehat{\theta} = (1 + |\widebar{\calR}|) / ( 1 + |\widebar{\calS}^{\pre}| )$.
\end{enumerate}

PSP above provides a flexible approach to subject selection. 
To illustrate its effectiveness and power, we compare it with the conformal selection approach in \citet{jin2023selection}.
Let $v:\calX \times \{0,1\} \rightarrow \bbR$ be the score function defined in Section 2.5 of \citet{jin2023selection}, and the corresponding conformal $p$-values are constructed as $p_j^{\JC} = \frac{1}{1+n} \left( 1 + \sum_{i\in[n]} \bbI \left( v(X_{m+i}, \bbI(Y_{m+i} \in \calC) ) \leq v(X_j, 0) \right) \right)$ for $j\in[m]$, and the selection set is given by $\calS^{\JC} = \{ j\in[m]: p_j^{\JC} \leq p_{( \widehat{l}^{\JC} )}^{\JC} \}$ with 
$
\widehat{l}^{\JC} = \max \{ l \in [m]: p_{(l)}^{\JC} \leq \frac{ l \alpha }{ m } \}.
$
A key distinction between the procedures for determining $\widehat{l}$ and $\widehat{l}^{\JC}$ is the inclusion of $\widehat{\theta}$, which partially helps explain the error gap and power loss observed with the regression residual score in \citet{jin2023selection}.
This issue is addressed by the carefully constructed clipped score function that $v(x, 1 )$ is sufficiently large for all $x\in\calX$, leading to $p_j^{\JC} = \frac{1}{1+n} \left( 1 + \sum_{i\in[n]: Y_{m+i} \notin \calC} \bbI \left( v(X_{m+i}, 0) \leq v(X_j, 0) \right) \right)$.

We note that in this problem of subject selection, since subjects with $Y_j \notin \calC$ are of no interest, as discussed in Remark \ref{remark:preclass}, we can simply set $\texttt{Pre}(x) \equiv 1$ in PSP, i.e., $\calS^{\pre} = [m]$ and $\widebar{\calS}^{\pre} = [n]$, and then PSP recovers the procedure with clipped score function in \citet{jin2023selection}, which presents effective error control and the superior power in the numerical studies therein.

\subsection{Informative Prediction Set Construction}\label{sec:extend_informative}

Conformal inference provides prediction sets for target responses with coverage guarantee.
However, these sets can sometimes be conservative in practice.
\citet{gazin2025selecting} addresses the issue by proposing informative prediction set selection and construction with false coverage rate (FCR) control, and introduces the \texttt{InfoSP} and \texttt{InfoSCOP} methods.
In this section, we focus on the length-restricted prediction in classifications as an example to illustrate how to extend and incorporate the idea of PSP to perform large-scale informative prediction set construction.
We remark that although PSP can also be applied for other types of informative sets and regression problems as in \citet{gazin2025selecting}, a general formulation is required, which is beyond the scope of this article and will not be presented here in details.

Let $(X_{m+i},Y_{m+i}) \in \calX \times [K], i\in[n]$ be the labelled data and let $X_j\in \calX,j\in[m]$ be the covariates of interest. 
Define the set of informative length-restricted sets as $\calI = \{ \calC \subset [K]: |\calC| \leq L \}$ (e.g., $L = \lfloor K /2 \rfloor $), and the goal is to select a subset $\calS \subset [m]$ and construct prediction sets $\{ \calC_j \}_{ j \in \calS }$ such that $\calC_j \in \calI$ for all $j\in\calS$ and $\text{FCR} = \bbE \sum_{ j\in\calS } \bbI ( Y_j \notin \calC_j ) / (1\vee |\calS|) \leq \alpha$.
Note that when $L=1$, the task simplifies to classification with overall error rate control in \eqref{eq:FDR}, and related numerical results are shown in Section \ref{sec:simulation}.

For general $1 \leq L \leq K-1$, let $\mu(x) = (\mu_1(x),\cdots,\mu_K(x)): \calX \rightarrow \Delta_{K-1}$ be a pre-trained classifier that learns the posterior probabilities, where $\Delta_{K-1}$ denotes the $(K-1)$-dimensional simplex set, and denote by $\mu_{(1)}(x) \leq \cdots \leq \mu_{(K)}(x)$ the sorted values. 
Let $\texttt{Pre}: \calX \rightarrow \{ 0,1 \}$ be a pre-trained pre-selection rule.
Then, PSP can be applied as follows for informative prediction set construction:
\begin{enumerate}[nosep]
\item Pre-selection. $\calS^{\pre} = \{ j\in[m]: \texttt{Pre}(X_j) = 1 \}$ and $\widebar{\calS}^{\pre} = \{ i\in[n]: \texttt{Pre}(X_{m+i}) = 1 \}$.

\item Selective $p$-value construction.
$
p_j = \frac{ 1 }{ 1+|\widebar{\calR}| } \left( 1 + \sum_{ i\in \widebar{\calR} } \bbI \left( s(X_{m+i}) \geq s(X_{j}) \right) \right)
$ for $j\in\calS^{\pre}$, where score function $s(x) = 1 - \mu_{(K-L)}(x)$, $\widebar{\calR} = \{ i\in \widebar{\calS}^{\pre} : Y_{m+i} \notin \calC_{m+i} = \calC(X_{m+i}) \}$, and prediction set $\calC(x) = \{ k\in[K]: \mu_k(x) > \mu_{ (K-L) } (x) \}$.

\item Post-selection inference: selection set $\calS = \{ j\in\calS^{\pre}: p_j \leq p_{(\widehat{l})} \}$ and prediction sets $\{ \calC_{j} = \calC(X_j) \}_{ j\in\calS }$, where 
$
\widehat{l} = \max \{ l \in [|\calS^{\pre}|]: p_{(l)} \leq \frac{ l \alpha }{ \widehat{\theta} |\calS^{\pre}| } \}
$
and $\widehat{\theta} = (1 + |\widebar{\calR}|) / ( 1 + |\widebar{\calS}^{\pre}| )$.
\end{enumerate}
Again, since selecting and constructing informative prediction sets is the only objective, we can set $\texttt{Pre}(x) \equiv 1$ and then $\calS^{\pre} = [m]$ and $\widebar{\calS}^{\pre} = [n]$.
Therefore, PSP provides a flexible way for the selective informative set construction.
Numerical results in Section \ref{sec:add_info} of the supplement demonstrate both the validity and power advantage of PSP.

\bibliographystyle{apalike}
\bibliography{ref}

\end{document}